\documentclass{ws-ijmpa}
\usepackage[super,compress]{cite}
\usepackage{graphicx}
\usepackage{multirow}
\begin{document}
\markboth{Zhiyang Yuan, Huirong Qi}{Feasibility study of TPC detector at high luminosity Z pole on the circular collider}

%
\catchline{}{}{}{}{}
%

\title{Feasibility study of TPC detector at high luminosity $Z$ pole on the circular collider}

\author{Zhiyang Yuan, Huirong Qi\footnote{Huirong Qi, qihr@ihep.ac.cn}, Yue Chang, Ye Wu, Hongyu Zhang, Jian Zhang, Yuanbo Chen}

\address{State Key Laboratory of Particle Detection and Electronics,\\
Institute of High Energy Physics, Chinese Academy of Sciences,\\
19B Yuquan Road, Beijing 100049, China\\
University of Chinese Academy of Sciences,\\
19A Yuquan Road, Beijing 100049, China}

\author{Yiming Cai, Yulan Li, Zhi Deng, Hui Gong}

\address{Department of Engineering Physics, Tsinghua University,\\
Hai Dian District, Beijing 100084, China}

\maketitle

\begin{history}
\received{Day Month Year}
\revised{Day Month Year}
\end{history}

\begin{abstract}
With the development of the circular collider, it is necessary to make accurate physics experimental measurements of particle properties at higher luminosity $Z$ pole. Micro-pattern gaseous detectors (MPGDs), which contain Gaseous Electron Multiplier (GEM) and Micro-mesh gaseous structures (Micromegas), have excellent potential for development as the readout devices of the time projection chamber (TPC) tracker detector. To meet the updated physics requirements of the high luminosity $Z$ from the preliminary concept design report (preCDR) to concept design report (CDR) at the circular electron positron collider (CEPC), In this paper, the space charge distortion of the TPC detector is simulated with the CEPC beam structure. Using the multi-physics simulation software package, the distribution of ion estimated by Geant4 is used as the input for the differential equation, and the relationship between the ion density distribution and electric field in the detector chamber is simulated. These simulation results show that the maximum deviation for Higgs $\mathcal{O}$(25~$\mu$m) meets the performance requirements in CEPC TPC detector at the high luminosity $Z$ pole, while it is still a considerable challenge for $Z$ pole, with the maximum deviation $\mathcal{O}$($\textgreater 100~\mu$m). According to the previous developments, the cascaded structure of GEM and Micromegas detector has been measured. The new considerations of the detector's requirements were given, the gain needs to be reached to about 2000 with IBF$\times$Gain under 0.1, and IBF means the ions back flow ratio of the detector. The pixel TPC is a potential option to replace the traditional MPGDs with the low gain, low occupancy, and outstanding pattern recognition. Finally, some update parameters and experiments results were compared.

\end{abstract}

\ccode{PACS numbers:~29.40.Cs}


\section{Introduction}
\label{sec:intro}
As one option of the CEPC track detector concept, TPC can meet the requirements of three-dimensional track reconstruction and charged particle identification, and has the advantages of the low material budget, outstanding pattern recognition capability and high position resolution. As the popularity and complexity of TPC applications in particle physics increases, its calibration and monitoring become important too. For the field distortion, although most sources of distortion are static, space charge distortion is a dynamic process due to the accumulation of positive ions in the TPC chamber. Most of the positive ions in the chamber are derived from the avalanche multiplication of electron clusters in the TPC readout structure. Under the action of an electric field, positive ions move slowly due to the drift velocity ($\mathcal{O}$(5~m/s)) compared to electrons ($\mathcal{O}$(10~m/ms)). The change in the type and quantity of charged particles entering the TPC chamber affects the distribution of positive ions in the chamber, and the electric field distortion generated by this distribution is eliminated until the cathode absorbs them. Continuous positive ion accumulation can indirectly cause changes in the transport properties of subsequent charged particles in the chamber\cite{a}. The CEPC baseline design uses a similar international linear collider detector (ILD) design concept. However, its detectors need to operate in continuous mode. For the TPC detector to meet the requirements of high position resolution and three-dimensional track reconstruction under continuous high count rate operation, correspondingly, the low IBF and sufficient cooling capacity are required. TPC readouts use cascaded MPGDs, especially GEM\cite{b} and Micromegas\cite{c} structures, which achieve very low IBF compared to original MWPCs readout structures\cite{d}.

\begin{figure}[htbp]
	\centering 
	\includegraphics[width=0.9\textwidth]{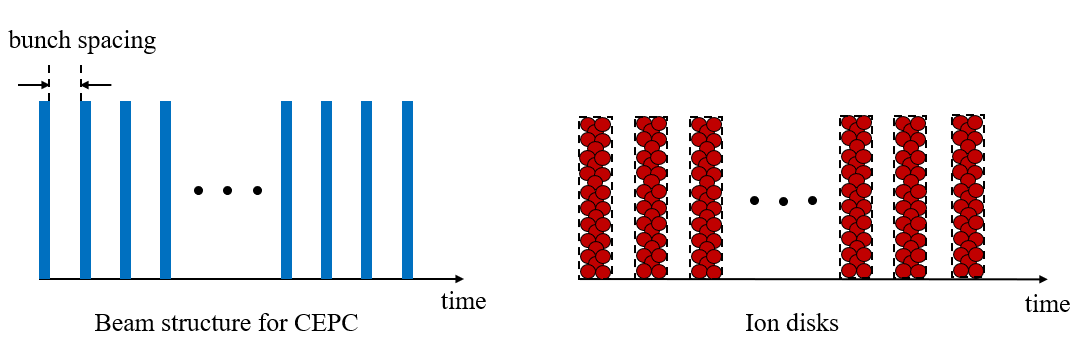}
	\caption{\label{fig:beam} The beam structure of CEPC with the specific bunch spacing for the Higgs, W and Z bosons (Left) and the structure of ion disk with equal spacing and thickness (Right).}
\end{figure}

The CEPC is a circular electron-positron collider with a 100~km circumference and two interaction points (IP). In the CEPC CDR\cite{e}, the bunch spacing as Fig.~\ref{fig:beam} (Left) of the Higgs, W and Z is approximately 760~ns, 200 $ns$ and 25 $ns$, respectively. On the one hand, the CEPC beam structure will cause the subsequent ions disk as Fig.~\ref{fig:beam} (Right). The longitudinal width of the ions disk is mainly affected by the bunch spacing. On the other hand, the luminosity for the Higgs, W and Z boson will strongly affect the ions density. To eliminate space charge effects in the CEPC TPC, we adopt a hybrid structure cascaded GEM with a Micromegas detector (GEM-MM) to suppress the back flow ions from the amplification process passively. Similarly, a structure cascaded double micro-mesh with Micromegas detector (DMM)\cite{f} has the same suppress function on IBF. Comparing to the traditional MPGDs, the pixel TPC\cite{g} combining a pixel ASIC with a MPGD is also an ideal selection. To evaluate the inhibitory effect, we need to simulate the space charge effects of primary and backflow ions. Conversely, the ILD TPC adopts a gate structure nearby the amplification to stop the ion disk entering the drift region actively. In the CEPC CDR, the luminosity for Higgs, W and Z bosons are $3\times 10^{34} ~$cm$^{-2}$s$^{-1}$, $10\times 10^{34} ~$cm$^{-2}$s$^{-1}$ and $32\times 10^{34} ~$cm$^{-2}$s$^{-1}$ respectively. Compared with the luminosity reported in the CDR for the CEPC, those of the ILC\cite{h} and FCC\cite{i} are $1.8\times 10^{34} ~$cm$^{-2}$s$^{-1}$ and $100\times 10^{34} ~$cm$^{-2}$s$^{-1}$.

In the following section, the space charge distortion of CEPC TPC is simulated and explained. In section 2, the theoretical model for space charge distortion is defined, the simulation results of $\phi$ deviation are presented, and some explanations are given. In section 3, the conclusion is finally reached.

\section{Feasibility Study Of The Tacker Module}
\subsection{Motivation and physics requirements}

The CEPC CDR and updated parameters are listed in the Table~\ref{tab:parameters}. The luminosity increase factor for Higgs and Z are $1.8$ and $3.2$ respectively from CEPC CDR to Updated parameters. Under the high luminosity operation, CEPC will produce close to one trillion Z bosons, 100 millions W boson and over one million Higgs to provide precision measurements of their properties. Additionally, it is an excellent opportunity to search for Beyond the Standard Model (BSM) physics. Correspondingly, the TPC as a tracker detector needs to provide perfect position resolution and a high count rate. The main problem is evaluating and solving space charge effects and distortions caused by the positive ions. The following subsections will give brief details about the evaluation of space charge effects in the CEPC TPC.

\begin{table}[htbp]
	\tbl{Updated parameters of collider ring since CDR.}
	{\begin{tabular}{lcccc}
		\toprule
		\multirow{2}*{Parameter}& \multicolumn{2}{c}{Higgs}&\multicolumn{2}{c}{Z} \\
		\cline{2-5}
		& CDR & Updated & CDR & Updated\\
		\colrule
		Beam energy ($GeV$) & 120 & - & 45.5 & -\\
		Synchrotron radiation loss/turn ($GeV$)& 1.73 & 1.68 & 0.036 & -\\
		Piwinski angle & 2.58 & 3.78 & 23.8 & 33\\
		Number of particles/bunch $N_{e} (10^{10})$ & 15.0 & 17 & 8.0 & 15\\
		Bunch number (bunch spacing) & 242 (0.68$\mu$s) & 218 (0.68$\mu$s) & 12000 & 15000\\
		Beam current ($mA$) & 17.4 & 17.8 & 461.0 & 1081.4\\
		Synchrotron radiation power /beam ($MW$) & 30 & - & 16.5 & 38.6\\
		Cell number/cavity & 2 & - & 2 & 1\\
		$\beta$ function at IP $\beta_{x}^{*}/\beta_{y}^{*}$ ($m$) & 0.36/0.0015 & 0.33/0.001 & 0.2/0.001 & -\\
		Emittance $\varepsilon_{x}/\varepsilon_{y}$ ($nm$) & 1.21/0.0031 & 0.89/0.0018 & 0.18/0.0016 & -\\
		Beam size at IP $\sigma_{x}/\sigma_{y}$ ($\mu$m) & 20.9/0.068 & 17.1/0.042 & 6.0/0.04 & -\\
		Bunch length $\sigma_{z}$ (mm) & 3.26 & 3.93 & 8.5 & 11.8\\
		Lifetime ($hour$) & 0.67 & 0.22 & 2.1 & 1.8\\
		Luminosity/IP L ($10^{34}$ cm$^{-2}$s$^{-1}$) & 2.93 & 5.2 & 32.1 &  101.6\\
		\botrule
	\end{tabular} \label{tab:parameters}}
\end{table}

\subsection{Simulation principles}
To study the interaction between the electric field $E$ and the ion density $n_{e, i^+}$ in the detector, the multi-physics software COSMOL was used to simulate the TPC geometry\cite{j,k}. It is a finite element analysis software, which can study transient and analyse the complex interaction process. In the detector geometry, the software can be used to calculate the effect of the ion density $n_{i^+}$ on the electric field. Throughout the physical process of the detector, COMSOL gives a numerical approximation of differential equations as the following:

\newcommand{\myvec}[1]%
{\stackrel{\raisebox{-2pt}[0pt][0pt]{\small$\rightharpoonup$}}{#1}}

\begin{equation}\label{eq:vector1}
	\myvec{\nabla }\cdot \myvec{E} = \frac{\rho }{\varepsilon },
\end{equation}

\begin{equation}\label{eq:vector2}
\myvec{n}\cdot (\myvec{D}_2-\myvec{D}_1) = \sigma,
\end{equation}

\begin{equation}\label{eq:vector3}
\rho = (1+k)\frac{L}{V_{ion}/(ms^{-1})}(\frac{2.74\times 10^{-3}}{r/mm-97.9}-1.25\times 10^{-6}) [fC/mm^{3}],
\end{equation}

\noindent where $\myvec{E}$ is the electric field, $\varepsilon$ is the dielectric constant, $\myvec{D}$ is the electric displacement vector, k is the ion backflow ratio factor (IBF$\times$Gain), L is the luminosity normalized to $1\times 10^{34}$cm$^{-2}$s$^{-1}$, $V_{ion}$ is the velocity of ions and r is the radius. E$_q$.~(\ref{eq:vector1}) is the Poisson equation which describes the potential field due to a given charge distribution. The presence of a large number of ions in the TPC causes the redistribution of electric charge in the outer shell, known as electrostatic induction. The charge on the surface of the outer shell is called induced charge, and its surface charge density $\sigma$ is described in E$_q$.~(\ref{eq:vector2}). The change in the normal component of the electric displacement vector is equal to the surface charge density. In this paper, the space charge density $\rho$ as E$_q$.~(\ref{eq:vector3})\cite{l}. Nine thousands $Z\to q\bar{q}$ events are simulated with Mokka, the Geant4 simulation package. The distribution of the ion density is shown in Fig.~\ref{fig:data} (Left) in CEPC TPC under the luminosity for Higgs. Considering the CEPC TPC is a columnar structure with an inner and outer radius of 0.3~m, 1.8~m and a full length of 4.7~m, the Poisson equation can be specifically written in the form of components with boundary conditions,the 3D geometry is approximated by 2D axial symmetry, and the axis of symmetry is the Z-axis. The electric field $\myvec{E}$ and magnetic field $\myvec{B}$ are parallel to the Z-axis. In the CEPC TPC, the drift chamber is filled with T2K gas~(Ar/CF$_4$/iC$_4$H$_{10}$=95/3/2). To reach a suitable drift velocity in the drift region, TPC adopts that the central cathode plane is at a high potential of 50~kV, and the two anodes at the two end-plates with the readouts are at ground potential.

\subsection{Electric field distortion and $\phi$ deviation}
\begin{figure}[htbp]
	\centering 
	\includegraphics[width=.45\textwidth]{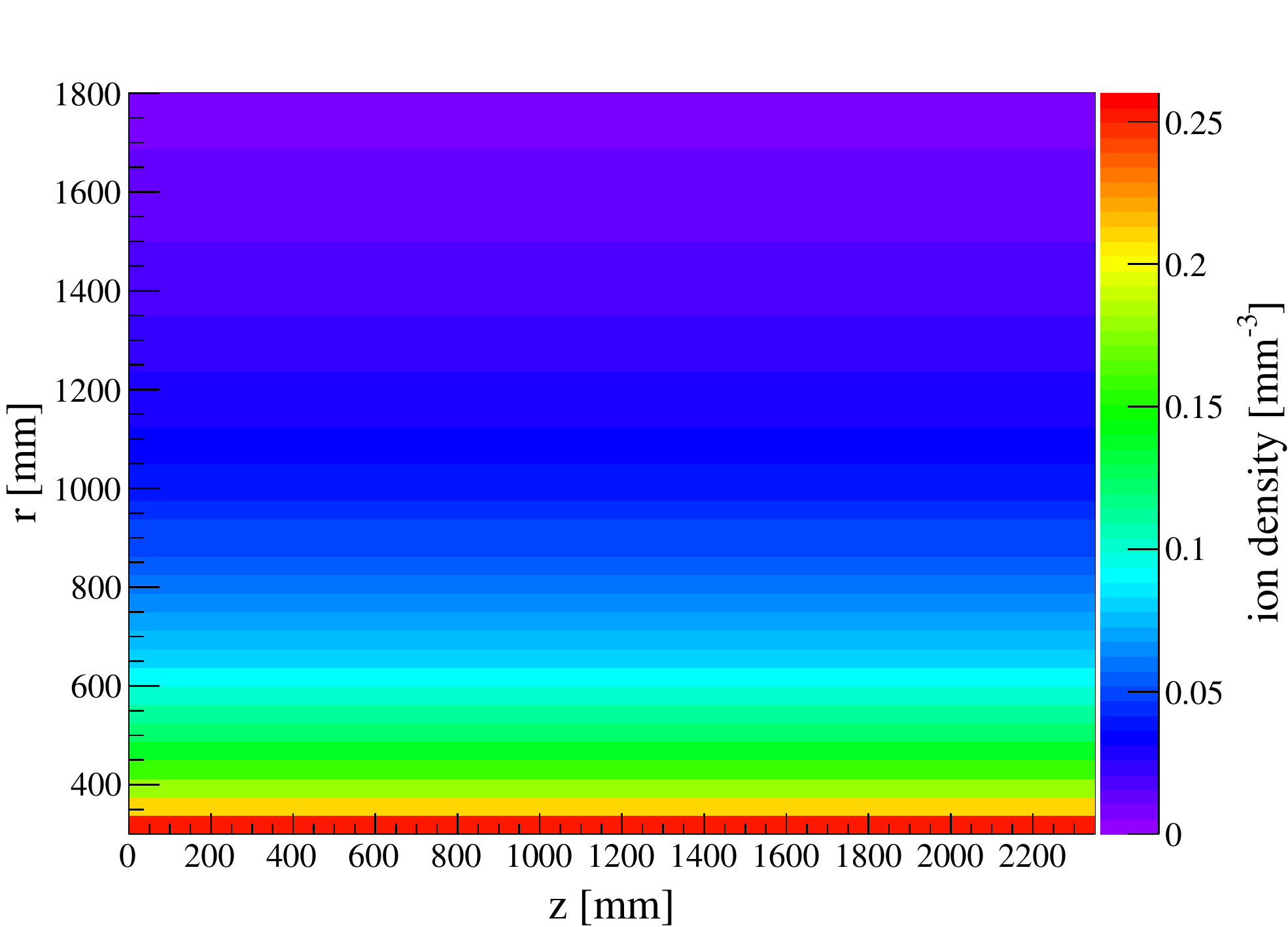}
	\qquad
	\includegraphics[width=.45\textwidth]{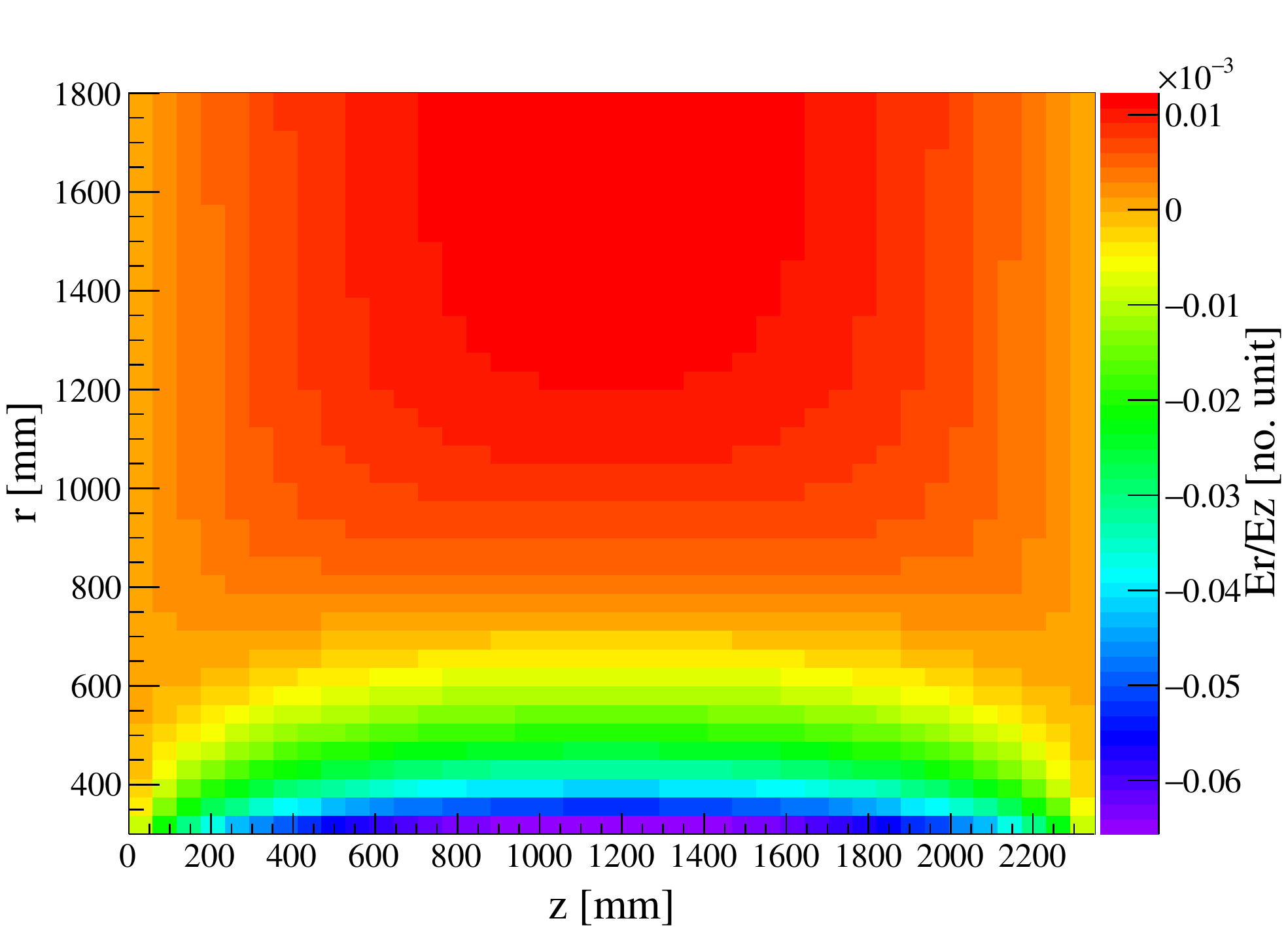}
	\caption{\label{fig:data} The distribution of the ion density (Left) and the value of $E_{r}/E_{z}$ (Right) in CEPC TPC under the luminosity for Higgs.}
\end{figure}

The motion of charged particles in an electromagnetic field calculated using Maxwell's equations. Considering that the electric field and the magnetic field are parallel, the relationship can be easily achieved between the moving distance in the z and $\phi$ direction. The small change of $\phi$ deviation $\Delta{D_{\phi}}$ is considered to be due to the small moving distance $\Delta{Z}$ in the z-direction and calculated by the E$_q$.~(\ref{eq:deviation}).

\begin{equation}\label{eq:deviation}
\Delta{D_{\phi}} = - \frac{\omega \tau}{1 + (\omega \tau)^2} \times \frac{E_{r}}{E_{z}} \times \Delta{Z},
\end{equation}

\noindent where $\omega \equiv eB/m$, $\tau$ is the mean free time of electrons, and E are are the $r$ and $z$ components of the electric field. The value of $\omega \tau$ equals 10 using T2K gas and the value of $E_{r}/E_{z}$ is shown in Fig.~\ref{fig:data} (Right).

\begin{figure}[htbp]
	\centering 
	\includegraphics[height=41mm,width=.45\textwidth]{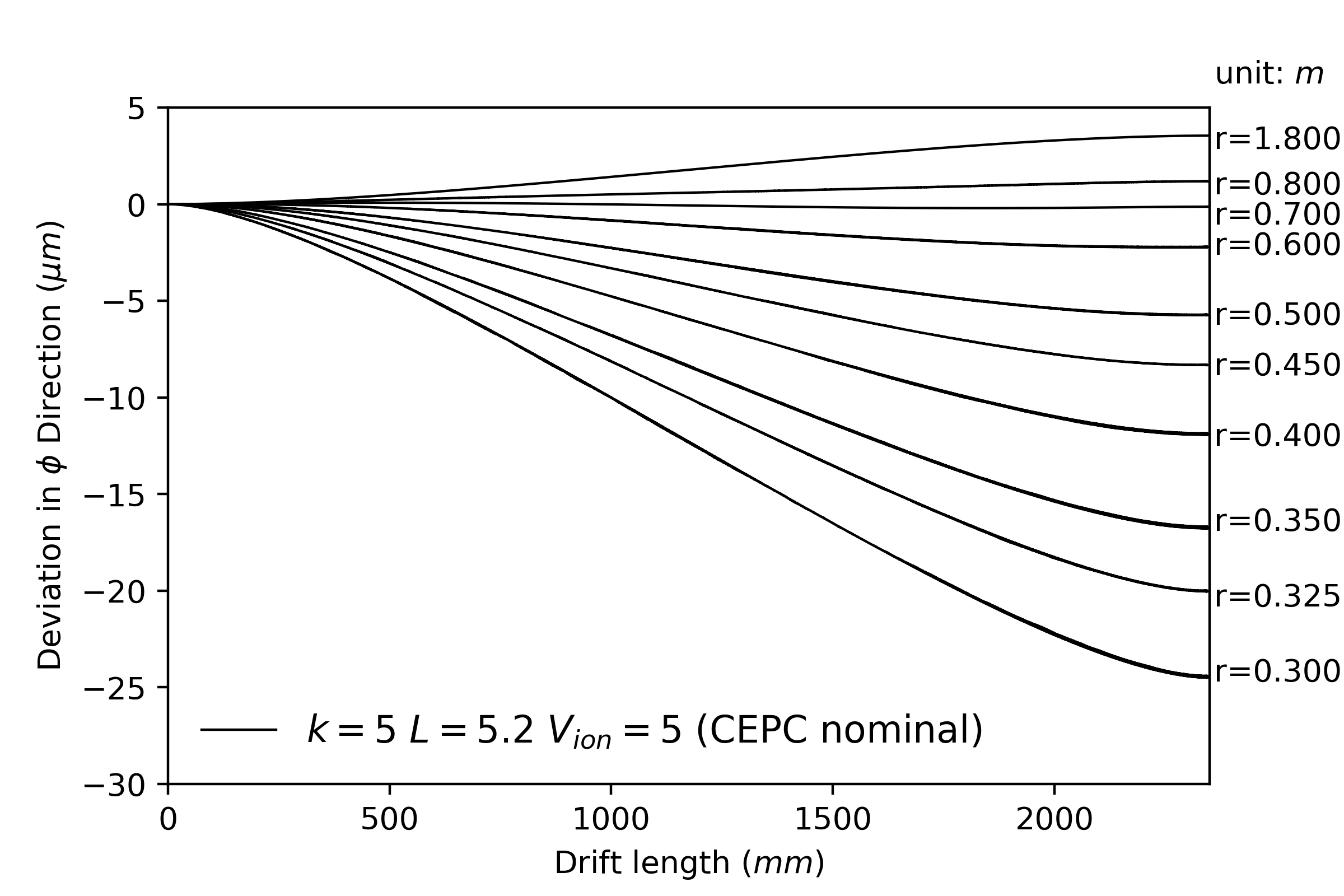}
	\qquad
	\includegraphics[height=41mm,width=.45\textwidth]{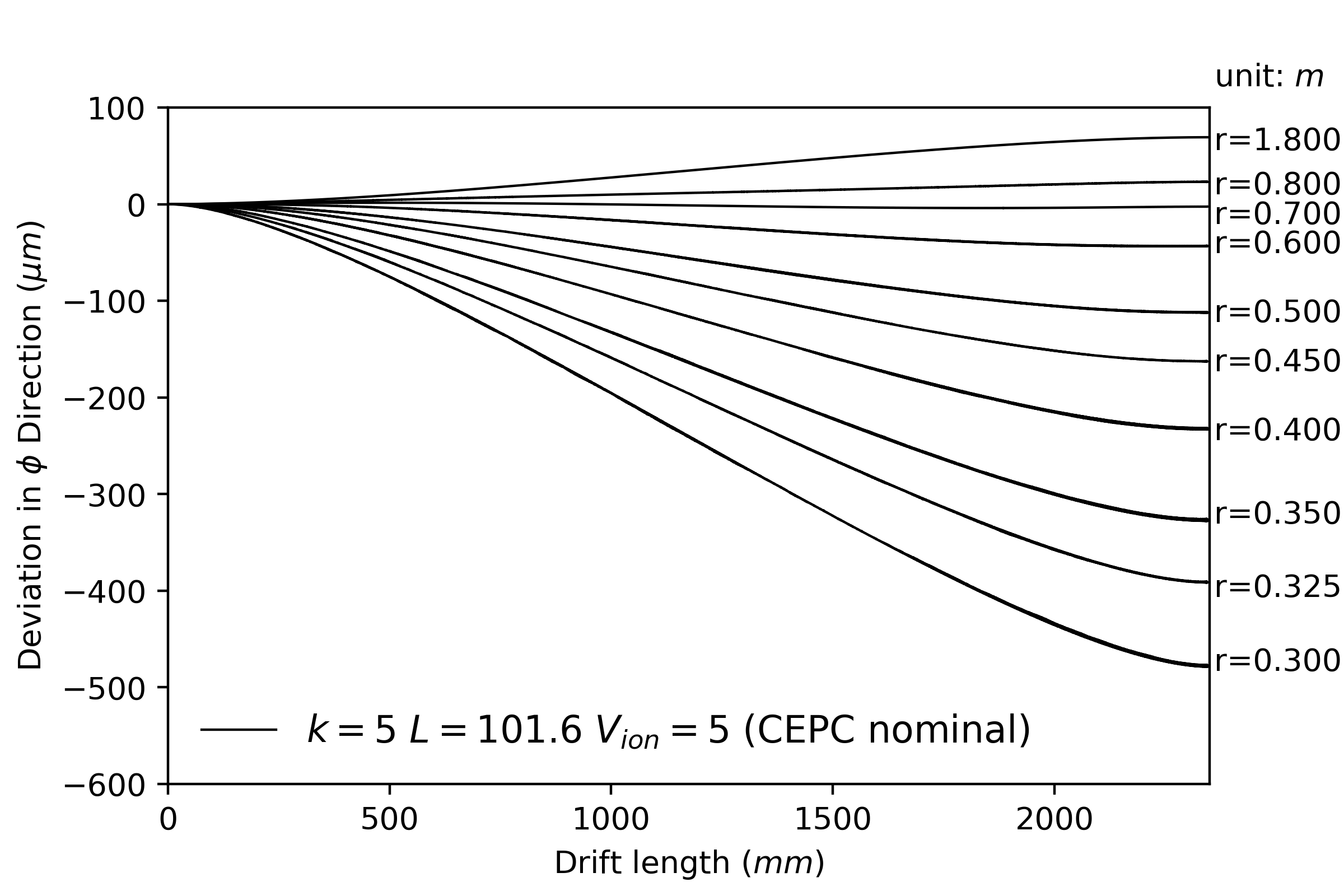}
	\caption{\label{fig:data1} The deviation in $\phi$ direction as a function of drift length with different luminosity for Higgs (Left) and Z (Right).}
\end{figure}

The deviation in the $\phi$ direction is shown in Fig.~\ref{fig:data1}. The simulation results consider the variation of radius and different luminosity for Higgs (Left) and Z (Right). Considering that the induced charge of the TPC cylinder will produce E$_r^{'}$, contrary to the direction of E$_r$ (offset effect), the opposite positional deviation will occur in the positive half axis. The E$_{r}$ includes the electric field caused by positive ion E$_{i^+}$ and the induced charge E$_r^{'}$.

\subsection{Related experimental research}
IBF research on MPGD has been tested and optimized at the Institute of High Energy Physics (IHEP) and the University of Science and Technology (USTC) of China. Experimental results of IHEP are carried out with mixture gases of T2K and Ar/iC$_4$H$_{10}$=95/5 separately and show that the IBF rate of GEM-MM can be reduced to about 0.1$\%$ at the gain of about 5000. Similarly, results of USTC are carried out with different tilt angles between the double meshes and show that the IBF rate of DMM can be as low as about 0.03$\%$ at a gain of about 20000. 
\begin{figure}[htbp]
	\centering 
	\includegraphics[height=40.5mm,width=.46\textwidth]{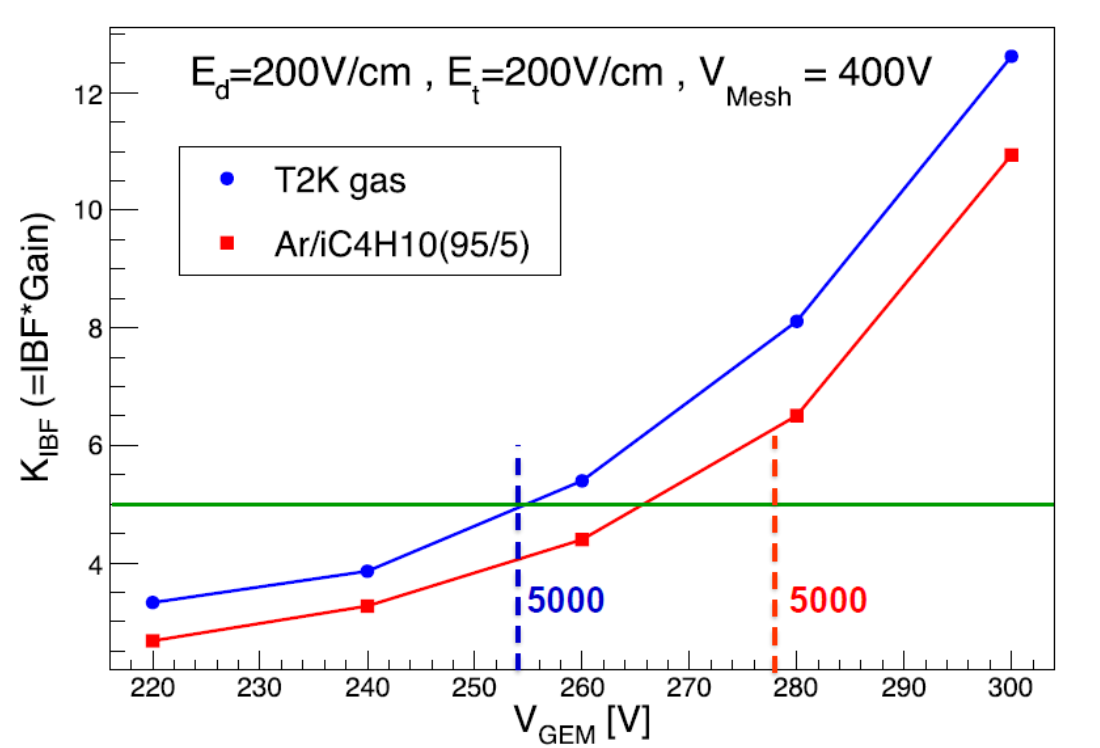}
	\qquad
	\includegraphics[height=40mm,width=.45\textwidth]{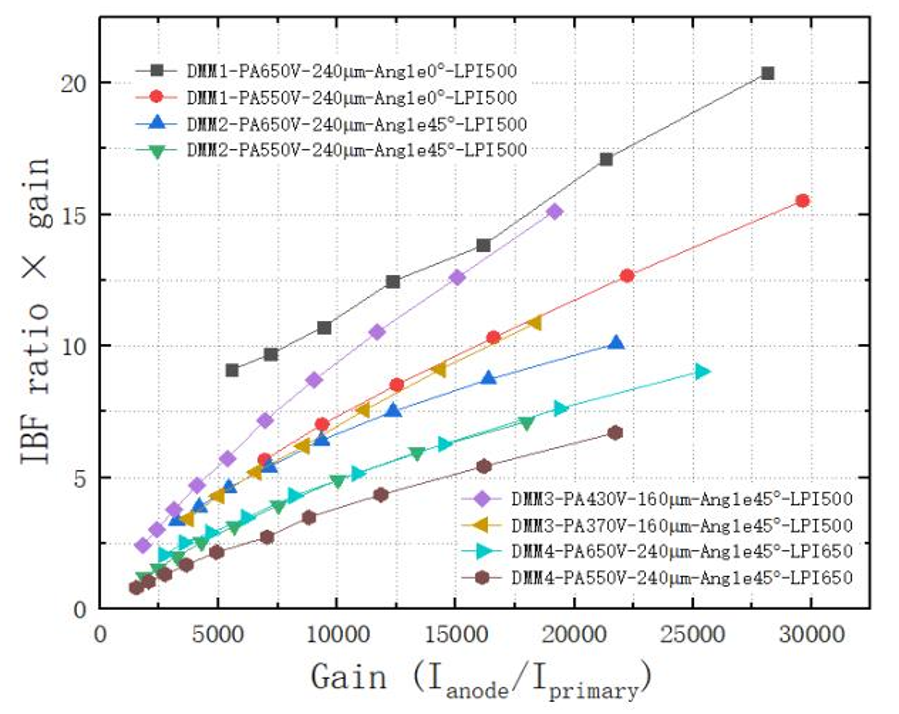}
	\caption{\label{fig:data2} The IBF$\times$Gain as a function of $V_{GEM}$ for GEM-MM\cite{m} (Left) and the IBF ratio as a function of gain with different tilt angles between the double meshes of a DMM\cite{f} (Right).}
\end{figure}

\section{Conclusion}
In this paper, we have simulated the $D_{\phi}$ of CEPC TPC with different luminosity for Higgs and Z. The maximum $\phi$ deviation for Higgs and Z is under 25~$\mu$m and 400~$\mu$m respectively. For Higgs luminosity, TPC only needs to ensure that the IBF$\times$Gain is less than 2 to achieve a less deviation of 10~$\mu$m. The deviation will be negligible enough to meet the performance requirements of CEPC TPC under $\mathcal{O}$(100~$\mu$m). However, it is a considerable challenge for Z with the similar physics requirements. The gain needs to be small enough ($<$2000) with IBF$\times$Gain under 0.1.

Nevertheless, IBF$\times$Gain has the limitation ratio from the detector R$\&$D at high gain. The new idea for pixel TPC is being considered as the option to take the place of the traditional MPGDs. Its gain is less than 2000, and there is almost no IBF$\times$Gain. It can handle the massive data rates during CEPC Z running. The pixel occupancies are low, and the pattern recognition will have no problem separating events and finding tracks. If CEPC produces close to one trillion or not one million Z bosons, the technology of TPC needs to be adopted. Moreover, the pixel TPC needs to be considered under a higher luminosity.

\section*{Acknowledgments}

The author thanks for Prof. Yuanning Gao, Prof. Yulan Li and Dr. Yiming Cai for some discussions of details. This study was supported by the National Natural Science Foundation of China (Grant NO.: 11975256), the National Key Programme for $S\&T$ Research and Development (Grant NO.: 2016YFA0400400) and the National Natural Science Foundation of China (Grant NO.: 11775242)


\end{document}